\documentclass[10pt,twoside]{article}
\usepackage{amssymb}
\usepackage{amsmath}
\usepackage{graphicx}
\usepackage[latin1]{inputenc}

\begin{document}

\title{Unitarity constraints for Yukawa couplings in the two Higgs doublet
model\ type III}
\author{Andr\'es Castillo\thanks{%
afcastillor@unal.edu.co}, Rodolfo A. Diaz\thanks{%
radiazs@unal.edu.co}, John Morales\thanks{%
jmoralesa@unal.edu.co} \\
Universidad Nacional de Colombia,\\
Departamento de F\'isica. Bogot\'a, Colombia.}
\maketitle

\begin{abstract}
Unitarity constraints for Yukawa couplings are considered in the Two Higgs
Doublet Model type III, by using a general expansion in partial waves for
fermionic scattering processes. Constraints over general Flavor Changing
Neutral Currents are found from that systematic, such bounds compete with
those coming from Lagrangian perturbativity requirement but are weaker than
those imposed from phenomenological processes and precision tests.
Nevertheless, for bounds based on unitarity the number of assumptions is the
lowest among phenomenological and theoretical limits. Indeed, these new
theoretical constraints are independent of scalar masses or mixing angles
for this extended Higgs sector, making them less model dependent.
\end{abstract}


\section{Introduction and 2HDM-Motivations}

The successful discovery of a Standard Model (SM)-like Higgs boson in the
mass region around 125-126 GeV \footnote{%
Recent analyses show that the data were consistent with a scalar with even
parity at 97$\%$ confidence level. The mass of the new particle has also
been measured precisely in ZZ decay channel: $m_{h}=125.8GeV$ \cite%
{newresults}.} could open a new stage of motivations in phenomenological
studies of Extended Higgs Sectors (EHS). The Two Higgs Doublet Model (2HDM)
is one of the simplest examples of EHS, which represents the non-trivial
next extension compatible with the gauge invariance and with the Electroweak
Spontaneous Symmetry Breaking (EWSSB) in the SM\footnotetext{%
The parameter $\rho \equiv m_{w}^{2}/m_{z}^{2}\cos ^{2}\theta _{w}$ is a
critical piece sensitive to the scalar structure of EWSSB. A very good
feature of models with extra $SU(2)$ doublets (or singlets) is that they do
not break the custodial $SU(2)$ global symmetry, that protects $\rho =1$ at
the tree level.}. In this framework, a second Higgs doublet is added with
the same quantum numbers as the first one. The 2HDM induces several
features, among them we have \cite{Diaz,Hunters,Sher}: i) the possibility of
generating either spontaneous or explicit CP violation, ii) the generation
of Flavor Changing Neutral Currents (FCNC) (inspired on neutrino
oscillations) still compatible with the strong experimental limits on them,
and iii) the prospect that some theories with larger symmetries end up at
low energies in a non minimal Higgs sector as in the case of SUSY and
Left-Right models.

The discrete symmetry $Z_{2}$ (e.g. $\Phi _{1}\leftrightarrow \Phi _{1}$ and 
$\Phi _{2}\leftrightarrow -\Phi _{2}$) is usually implemented in the 2HDM
because it forbids mixing between the two doublets. Moreover, this symmetry
ensures a CP- conserving frame in the scalar sector and FCNCs are forbidden.
In the last point, the $Z_{2}$ transformation unfolds the 2HDM since its
presence (absence) leads to different forms of the Yukawa couplings between
fermions and Higgs bosons with (without) flavor natural conservation.

In the CP-conserving frame, after the EWSSB all eight components of the two
doublets are transformed into three Goldstone bosons ($G^{0},G^{\pm }$), two
CP-even Higgs bosons ($h^{0},H^{0}$), one CP-odd Higgs boson ($A^{0}$) and
two charged Higgs bosons ($H^{\pm }$). Although experimental accessibility
is now significant, those scenarios of 2HDM remain unknown. Hence,
theoretical constraints must be obtained by making additional fundamental
assumptions on the Quantum Field Theory background of the 2HDM. Among these
constraints we have: Vacuum stability, triviality, oblique parameters, and
the unitarity at the tree level. Triviality limits rely on the fact that the
quartic couplings in the potential remain finite up to large scales of
energy. On the other hand, the condition of vacuum stability requires that
couplings in the Higgs potential must be positive in all field space
directions for their asymptotically large values, otherwise the potential
would be unbounded from below and had no minimum.

On the other hand, $S-$matrix unitarity is embodied in all perturbative
levels due to the Optical Theorem (OT)\cite{Peskin}. Especially, for
asymptotically flat models (wherein scattering amplitudes do not exhibit any
power-like growth in the high energy limit), and main contributions lie at
the tree level. Therefore, the other theoretical constraints are guaranteed
through forthright realization of perturbation theory.

Perturbative unitarity bounds are derived from the Lee-Quigg-Thacker (LQT)
method which shows that if the Higgs boson mass $m_{\eta ^{0}}\ $exceeds
critical values obtained from partial wave decomposition, unitarity is
violated at the tree level for different binary scattering processes $%
p_{1}p_{2}\rightarrow p_{3}p_{4}$ at high energies, $s>>m_{\eta ^{0}}^{2}$ 
\cite{Dicus,LQT,Kanemura,Akeroyd,Kladiva}.

These constraints have only been applied to bosonic scalar and bosonic
vectorial sectors because the unitarity limits emerge from Plane Waves
Decomposition, which is only valid for spinless particles (for vector bosons
this formalism can be applied because of the theorem of equivalence). For
fermionic sectors it is necessary to introduce a General Partial Waves
Decomposition where different spin states are involved. The Jacob-Wick
expansion is the most natural and simplified method for partial
diagonalization in the angular momentum basis, since such expansion relies
on the appropriate choice of physical states in the initial characterization
of scattering processes. Attempts to constrain the fermionic sector of four
generation models with criteria of unitarity (but without this approach)
have been carried out in \cite{Dawsonunitarity}.

To incorporate all these concepts, we first show some few issues of the 2HDM
in section \ref{Fundamentals}, where the general Yukawa sector is introduced
and the FCNC's sources are described. At this point, we review the different
phenomenological methods to obtain bounds on Yukawa couplings. In section %
\ref{JWe}, we consider the general form of the Jacob-Wick approach.
Assumptions over helicity states are imposed in order to obtain a decoupling
regime of this method to the traditional wave expansion and thereby to build
up unitarity constraints over their associated coefficients. Then in section %
\ref{fermionic}, some particular scattering processes involving fermionic
interactions are studied within the framework of the 2HDM\ type III. The
last section \ref{charged} is dedicated to extrapolate this method to the
study of charged channels of fermionic interactions. Finally, some
concluding remarks are given, and some prospects for further studies
involving others models are established.

\section{Fundamentals of the 2HDM}

\label{Fundamentals}

The 2HDM is one of the simplest non-trivial extension of the Higgs sector in
the Standard Model, where two doublets of scalars with unit hypercharge are
considered in order to achieve the Electroweak Spontaneous Symmetry
Breaking. The full Lagrangian of the most general 2HDM contains additional
terms with respect to the minimal SM in the scalar potential, the kinetic
sector and the Yukawa interactions sector.

Since there are two vacuum expectation values (VEV), one for each doublet,
there could be in general a relative phase between them. This fact is the
source for a spontaneous CP-violation coming from the EWSSB, which produces
a vacuum different from a neutral one. Moreover, if the parameters in the
scalar and Yukawa potentials are complex, explicit CP violation also arises.
Since we are not going to treat these effects here, we assume all parameters
and VEVs to be real in the following. This formalism will be named the%
\textit{\ CP-conserving} \textit{2HDM frame}. On the other hand, the most
general Yukawa Lagrangian has a set of dimension-4 Higgs Fermion couplings,
given by

\begin{eqnarray}
-\mathcal{L}_{Y} =\eta _{i,j}^{U,0}\bar{Q}_{iL}^{0}\tilde{\Phi}%
_{1}U_{jR}^{0}+\eta _{i,j}^{D,0}\bar{Q}_{iL}^{0}{\Phi }_{1}D_{jR}^{0}+\xi
_{i,j}^{U,0}\bar{Q}_{iL}^{0}\tilde{\Phi}_{2}U_{jR}^{0}+\xi _{i,j}^{D,0}\bar{Q%
}_{iL}^{0}{\Phi }_{2}D_{jR}^{0}+\text{leptonic sector}+h.c.,  \label{LY}
\end{eqnarray}

where ${Q}_{iL}^{0}$ denote the left-handed quark doublets -with $i$ the
family index-. $U_{jR}$ ($D_{jR}$) correspond to the right-handed singlets
of up-type (down-type) quarks. $\eta _{ij}^{0}$ and $\xi _{ij}^{0}$ are
non-diagonal$\ 3\times 3\ $matrices. The superscript {}\textquotedblleft
0\textquotedblright\ indicates that the fields are not mass eigenstates yet.

The general Yukawa Lagrangian in (\ref{LY}) leads to processes with Flavour
Changing Neutral Currents (FCNC) at the tree level. It is due to the fact
that by rotating the down sector of quarks (or up and lepton sectors) to get
the mass eigenstates it is not possible to diagonalize both coupling
matrices $\eta _{ij}^{0},\xi _{ij}^{0}$ simultaneously.

Processes containing FCNCs are strongly supressed experimentally, in
particular due to the small $K_{L}-K_{s}$ mass difference. In SM the FCNC
are strongly suppressed by virtue of the GIM mechanism \cite{GIM}. In the
2HDM, several mechanisms to suppress FCNC at the tree level were proposed.
One of them is to consider the exchange of heavy scalar or pseudoscalar
Higgs Fields or by cancellation of large values with opposite sign. In 2HDM,
one mechanism is provided by Glashow and Weinberg, who implemented in the
Yukawa Lagrangian a discrete simmetry that automatically forbids the
couplings among fermions and scalars that generate such rare decays. From $K-%
\bar{{K}}$ mixing, as well as many processes involving kaon and muon decays 
\cite{McWilliamsLiShanker}, it has been considered that the heaviest fermion
set the scale for the entire matrix of Yukawa couplings. This assumption
yields many stringent bounds for the heavy scalars, for instance 150 TeV
(lower bound) from $K-\bar{{K}}$ mixing.

However the most outstanding feature of the fermion masses is their
hierarchical structure. If we expect roughly the same hierarchy in the
Yukawa couplings, setting all the Flavour Changing (FC) couplings to be of
the order of the heaviest-fermion Yukawa couplings is not reliable \cite%
{Sher}. From these considerations, Cheng and Sher proposed that FC couplings
should be of the order of the geometric mean of the Yukawa couplings of the
two fermions. Such an ansatz leads to a parametrization for the Yukawa
couplings of the form

\begin{equation}
\xi _{ij}=\frac{{\lambda _{ij}\sqrt{{2m_{i}m_{j}}}}}{v},  \label{cheng-sher}
\end{equation}%
because under the Cheng and Sher ansatz we expect that $\lambda _{ij}$ $\sim 
\mathcal{O}(1)$. If the Cheng and Sher ansatz is correct, the FCNC coming
from the first two generations are strongly supressed since the associated
Yukawa couplings are. As a consequence, the lower bound on Higgs boson
masses is reduced \cite{Sher,cheng}.

It is usual to assume the validity of the Cheng and Sher ansatz in the 2HDM
type III, and its implications are explored in different studies. Many
searches focus on some few specific processes, including $\Delta m_{B},%
\hspace{1em}{t\rightarrow ch}$ and $h\rightarrow \bar{{t}c}+\bar{{c}t}$,
rare $\mu ,\tau ,$ and $B$ decays ($B\rightarrow K\mu \tau $), $\mu
\rightarrow e\gamma $ at the two loop level, $t\rightarrow c\gamma $ and $%
t\rightarrow cZ^{0}$, muon-electron conversion, and $b\rightarrow s\gamma $.

If the Cheng and Sher ansatz is correct, then one would expect from $\lambda
_{ij}$ to be all of order unity (emulating to SM). This request is very weak
since there are unknown mixing angles. In addition, for many
phenomenological limits, several scalar masses enter in all specific
processes.

One of the most stringent experimental constraints on the 2HDM comes from
flavor physics. For the 2HDM with general (flavor diagonal) Yukawa
couplings, and from the charged Higgs contribution to different transitions
such as $b\rightarrow s\gamma $ in \cite{Mahmoundistal}, bounds on $|\lambda
_{tt}|$ were found; which must be less or equal to unity when $m_{H^{\pm
}}\lesssim 500$ GeV. With theoretical and experimental assumptions over SM
predictions (from QCD lattice) of observables involving $F-\bar{F}$ mixing
(with $F\equiv K,D,B_{d}$ or $B_{s}$) and with the measured meson mass
differences $\Delta M_{F}$ \cite{5Gupta}, contraints over flavor space have
been computed. Reference \cite{Gupta} considers that the addition between SM
and the new contribution does not exceed the experimental values, $\Delta
M_{B_{d}}^{expt}=(3.337\pm 0.033)\times 10^{-13}$ GeV and $\Delta
M_{B_{s}}^{expt}=(117.0\pm 0.8)\times 10^{-13}$ GeV, by more than two
standard deviations for $B_{d}$ and $B_{s}$ systems. This procedure impose
upper bounds on $\xi _{db}$ and $\xi _{sb}$ Yukawa couplings. Moreover for $%
K $ and $D$ systems, in order to obtain the upper bounds on $\xi _{ds}$ and $%
\xi _{uc}$ it is required that only the 2HDM contribution does not exceed
the experimetal values, $\Delta M_{K}^{expt}=(3.476\pm 0.006)\times 10^{-15}$
GeV and $\Delta M_{D}^{expt}=(0.95\pm 0.37)\times 10^{-14}$ GeV, by more
than two standard deviations. For a mass degenerate spectrum in neutral
scalar and pseudoscalar sector $m_{h^{0}}=m_{H^{0}}=m_{A^{0}}=120$ GeV those
bounds are transformed in limits on Cheng-Sher couplings: $(\lambda
_{ds},\lambda _{uc},\lambda _{bd},\lambda _{bs})\leq (0.1,0.2,0.06,0.06).$
Another phenomenological bounds on Cheng-Sher couplings for quark and
leptonic sectors are revised in table \ref{tab:bounds}.

\begin{table}[tbp]
\begin{center}
\begin{tabular}{|c|c|c|c|c|}
\hline
{$\lambda_{ij}$} & {Process} & {Assumptions} & {Bound} & Reference \\ 
\hline\hline
$\sqrt{\lambda_{bs}\lambda_{\mu\tau}}$ & $B\to K\mu\tau$ & Presicion Test & $%
\lesssim \mathcal{O}(10)$ & \cite{Aubert} \\ \hline
$\sqrt{\lambda_{ut}\lambda_{ct}}$ & $D-\bar{D}$ & $100\leq$scalar masses
(GeV)$\leq400$ & $\leq0.6$ & \cite{Golowich} \\ \hline
$\lambda_{tt}$ & $b\to s\gamma$ & $\lambda_{ii}=0$ $i\neq t,b$ and $%
m_{H^{\pm}}\leq300$ GeV & $\lesssim1.7$ & \cite{Ltt} \\ \hline
$\lambda_{\mu\tau}$ & $(g-2)_{\mu}$ & $m_{A^{0}}>>m_{H^{0}},m_{h^{0}}$ & $%
(10,80)$ & \cite{Diazart} \\ \hline
$\lambda_{e\tau}\lambda_{\mu\tau}$ & $(g-2)_{\mu}$ & $%
m_{A^{0}}>>m_{H^{0}},m_{h^{0}}$ & $<0.004$ & \cite{Diazart} \\ \hline
$\lambda_{e\tau}$ & $(g-2)_{\mu}$ & $m_{A^{0}}>>m_{H^{0}},m_{h^{0}}$ & $%
<10^{-3}$ & \cite{Diazart} \\ \hline
\end{tabular}%
\end{center}
\par
\label{tab:bounds}
\caption{\textit{Phenomenological Bounds on Yukawa couplings in the 2HDM
type III under Cheng and Sher anzats.}}
\end{table}

From this phenomenological review, it is clear that bounds on Yukawa
couplings depend strongly on the Higgs mass pattern (and also from other
free parameters such as the mixing angles). In what follows, we consider a
theoretical limit for the fermionic sector through general unitarity
constraints, where by means of a helicity formalism, it is possible to find
bounds on the general structure of Yukawa couplings. We shall see that,
despite those bounds are weaker than the phenomenological ones, such limits
are independent of the Higgs masses and mixing angles, making them more
model independent.

\section{An alternative approach to the unitarity constraints on Yukawa
couplings}

\label{JWe}

With the aim of obtaining unitarity constraints of scattering processes that
involves fermionic states, we introduce a general result based on an
expansion of partial waves through the so-called Jacob-Wick Formalism (JWF) 
\cite{Jacob}. The JWF is based on a diagonalization (at least partial) in
the angular momentum basis of the $\widehat{S}$ matrix for the scattering of
two particles (with initial helicities $\lambda _{a},\lambda _{b}$ and final
helicities $\lambda _{c},\lambda _{d}$) in the center of mass frame\footnote{%
In the helicity formalism, the spin degrees of freedom of the particle
involved do not introduce any significant complication with respect to
spinless particles. By contrast, in the conventional approach with static
spin labels for the particles, the relationship between {}\textquotedblleft
plane wave\textquotedblright\ and {}\textquotedblleft angular
momentum\textquotedblright\ states, leads to multiple Clebsh-Gordan
couplings coefficients for both the initial and the final states, and
consequently the partial wave is much more complicated than that of spinless
particles \cite{Wu-Ki-Tung}}. This procedure leads to the following general
expansion of the invariant amplitude \cite{Jacob,Wu-Ki-Tung,LibroHorejsi}

\begin{equation*}
\mathcal{M}(s,\Omega )=16\pi \sum_{J}(2J+1)\mathcal{D}_{(J)}^{\dagger
}(\Omega )_{\hspace{0.5cm}\lambda _{c}-\lambda _{d}}^{\lambda _{a}-\lambda
_{b}}\mathcal{M}^{J}(s),
\end{equation*}%
where $s$ is the CM energy, and $\Omega \equiv (\theta ,\phi )$ defines the
scattering polar and azimuthal angles. In addition, $\mathcal{D}(\Omega
)^{\lambda }{}_{\lambda ^{\prime }}$ are Wigner functions (from Rotations
Group Representations \cite{Wu-Ki-Tung}). These functions acomplish the
following orthogonality relation

\begin{equation}
\int \mathcal{D}_{m_{1}m_{1}^{\prime }}^{\ast (j_{1})}(\Omega )\mathcal{D}%
_{m_{2}m_{2}^{\prime }}^{(j_{2})}(\Omega )\frac{d\Omega }{4\pi }=\frac{1}{%
2j_{1}+1}\delta _{m_{1}m_{2}}\delta _{m_{1}^{\prime }m_{2}^{\prime }}\delta
_{j_{1}j_{2}}.  \label{ortho}
\end{equation}%
$\mathcal{M}^{J}(s)$ represents partial wave amplitudes obtained from the
matrix element $\mathcal{M}^{J}(s)\equiv \langle \lambda _{c}\lambda _{d}||%
\mathcal{M}(s,\Omega )||\lambda _{a}\lambda _{b}\rangle $. From (\ref{ortho}%
) we can get an explicit form for these coefficients

\begin{align}
\mathcal{M}^{J}(s)=\frac{1}{16\pi}\int\mathcal{M}(s,\Omega)\mathcal{D}%
^{*(J)}(\Omega)_{\hspace{0.5cm}\lambda_{c}-\lambda_{d}}^{\lambda_{a}-%
\lambda_{b}}d\Omega.
\end{align}

If the initial and the final helicity states are zero (i.e. $%
\lambda_{a}=\lambda_{b}$ and $\lambda_{c}=\lambda_{d}$), the $\mathcal{D}$
functions are reduced to Legendre polynomials,

\begin{align}
\mathcal{M}^{J}(s)=\frac{1}{32\pi}\int_{-1}^{1}\mathcal{M}%
(s,\theta)P^{J}(\cos\theta)d(\cos\theta).  \label{eq:onda}
\end{align}

This relation leads to the form of the coefficients in the traditional
partial waves decomposition. At the high energy limit, perturbative
unitarity requires that

\begin{align}  \label{uni}
\vert\text{Re}(\mathcal{M}^{J}(s))\vert\leq\frac{1}{2}.
\end{align}

\section{$f\bar{f}\to f\bar{f}$ processes}

\label{fermionic}

\begin{figure}[tph]
\centering\includegraphics[scale=0.3]{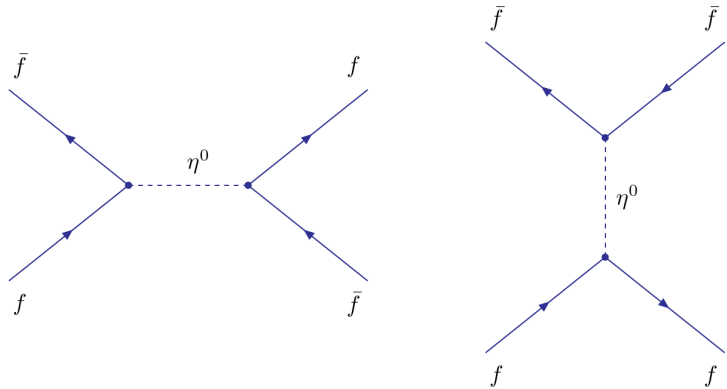} \vspace{0.25cm}
\caption{\textit{Diagrams at the tree level with scalar contribution
(neutral Higgs bosons) for $f\bar{f}\rightarrow f\bar{f}$ processes.}}
\label{fig:ff}
\end{figure}

In the following we consider the tree level matrix elements for the process $%
f\bar{f}\rightarrow f\bar{f}$ at the high energy limit under the helicity
spinors formalism. The neutral Higgs ($\eta ^{0}$) contributions are shown
in Fig. \ref{fig:ff}.


First, we write the invariant amplitude in the CP-conserving frame:

\begin{align}
\mathcal{M}(f\bar{f}\rightarrow f\bar{f})& =\sum_{\eta _{CP-even}^{0}}\left( 
\bar{v}_{2}i\chi _{ff}^{\eta ^{0}}u_{1}\frac{1}{s-m_{\eta ^{0}}^{2}}\bar{u}%
_{3}i\chi _{ff}^{\eta ^{0}}v_{4}+\bar{v}_{2}i\chi _{ff}^{\eta ^{0}}v_{4}%
\frac{1}{t-m_{\eta ^{0}}^{2}}\bar{u}_{3}i\chi _{ff}^{\eta ^{0}}u_{1}\right) 
\notag \\
& +\sum_{\phi _{CP-odd}^{0}}\left( \bar{v}_{2}i\chi _{ff}^{\phi ^{0}}\gamma
_{5}u_{1}\frac{1}{s-m_{\phi ^{0}}^{2}}\bar{u}_{3}i\chi _{ff}^{\phi
^{0}}\gamma _{5}v_{4}+\bar{v}_{2}i\chi _{ff}^{\phi ^{0}}\gamma _{5}v_{4}%
\frac{1}{t-m_{\phi ^{0}}^{2}}\bar{u}_{3}i\chi _{ff}^{\phi ^{0}}\gamma
_{5}u_{1}\right) ,
\end{align}%
here $\chi _{ff}^{\eta ^{0}}$($\chi _{ff}^{\phi ^{0}}$) are the couplings
between fermions and CP-even (CP-odd) neutral Higgs bosons. For different
helicity combinations that satisfy the relations $\lambda _{a}=\lambda _{b}\ 
$and $\lambda _{c}=\lambda _{d}$, we get the following non-zero coupled
channels (see appendix \ref{appendix})

\begin{subequations}
\begin{align}
\mathcal{M}(f_{\uparrow }\bar{f}_{\uparrow }\rightarrow f_{\uparrow }\bar{f}%
_{\uparrow })& =-\sqrt{2}G_{f}m_{f}^{2}\sum_{\eta _{all}^{0}}\left( \Xi
_{ff}^{\eta ^{0}}\right) ^{2}\left( \frac{s}{s-m_{\eta ^{0}}^{2}}\right)
\label{M1} \\
\mathcal{M}(f_{\downarrow }\bar{f}_{\downarrow }\rightarrow f_{\uparrow }%
\bar{f}_{\uparrow })& =-\sqrt{2}G_{f}m_{f}^{2}\sum_{\eta _{all}^{0}}\left(
\Xi _{ff}^{\eta ^{0}}\right) ^{2}\left( \frac{s}{s-m_{\eta ^{0}}^{2}}-\frac{t%
}{t-m_{\eta ^{0}}^{2}}\right) ,  \label{M2}
\end{align}%
where $s\ $and $t$ are Mandelstam variables and $\Xi _{s}$ are relative
Yukawa couplings of neutral Higgs bosons with respect to SM. The input
associated with the pseudoscalar sector is the same as to the scalar sector,
because in the high energy limit the eigenspinors are also chiral
eigenstates (see appendix \ref{appendix}).

There are other channels, e.g. $\mathcal{M}(f_{\downarrow }\bar{f}%
_{\downarrow }\rightarrow f_{\downarrow }\bar{f}_{\downarrow })$ and $%
\mathcal{M}(f_{\uparrow }\bar{f}_{\uparrow }\rightarrow f_{\downarrow }\bar{f%
}_{\downarrow })$, which just differ by a minus sign from amplitudes (\ref%
{M1}) and (\ref{M2}) respectively. Therefore, they would not provide new
information about unitarity bounds.

The $J=0$ partial wave coefficient is given by (\ref{eq:onda}). From the
definition of the Mandelstam variables, we have in the high energy limit

\end{subequations}
\begin{equation}
a^{0}\equiv \mathcal{M}^{J=0}(s)=\frac{1}{16\pi s}\int_{-s}^{0}\mathcal{M}%
(s,t)dt.  \label{coefi}
\end{equation}%
Therefore, in the regime of $s>>m_{\eta ^{0}}^{2}$ and for elastic
scattering channels (e.g. $f_{i}\bar{f_{i}}\rightarrow f_{i}\bar{{f_{i}}}$),
the non-zero matrix elements lead us to the following coefficients

\begin{eqnarray}
a^{0}(f_{\uparrow }\bar{f}_{\uparrow }\rightarrow f_{\uparrow }\bar{f}%
_{\uparrow }) &=&-\frac{{\sqrt{2}G_{f}m_{f}^{2}}}{16\pi }\sum_{\eta
_{all}^{0}}\left( \Xi _{ff}^{\eta ^{0}}\right) ^{2},  \label{a01'} \\
a^{0}(f_{\downarrow }\bar{f}_{\downarrow }\rightarrow f_{\uparrow }\bar{f}%
_{\uparrow }) &=&0.  \label{a0}
\end{eqnarray}

For the 2HDM type III (in the fundamental parametrization i.e. $\tan \beta
=0 $) the Yukawa couplings for neutral interactions are displayed in table %
\ref{tabla de acoples down I}.

\begin{table}[htp]
\centering
\par
\begin{center}
\begin{tabular}{|c|c|c|}
\hline
Coupling/Model & 2HDM III (up-sector) & 2HDM III (down-sector) \\ 
\hline\hline
$\Xi_{q_{i}\bar{q}_{j}}^{H^{0}}$ & $\left(\delta_{ij}\cos\alpha+\frac{%
\xi_{ij}^{U}\sin\alpha}{\sqrt{2m_{u_{i}}m_{u_{j}}}}v\right)$ & $%
\left(\delta_{ij}\cos\alpha+\frac{\xi_{ij}^{D}\sin\alpha}{\sqrt{%
2m_{d_{i}}m_{d_{j}}}}v\right)$ \\ \hline
$\Xi_{q_{i}\bar{q}_{j}}^{h^{0}}$ & $\left(-\delta_{ij}\sin\alpha+\frac{%
\xi_{ij}^{U}\cos\alpha}{\sqrt{2m_{u_{i}}m_{u_{j}}}}v\right)$ & $%
\left(-\delta_{ij}\sin\alpha+\frac{\xi_{ij}^{D}\cos\alpha}{\sqrt{%
2m_{d_{i}}m_{d_{j}}}}v\right)$ \\ \hline
$\Xi_{q_{i}\bar{q}_{j}}^{A^{0}}$ & $-i\frac{\xi_{ij}^{U}}{\sqrt{%
2m_{u_{i}}m_{u_{j}}}}v\gamma_{5}$ & $i\frac{\xi_{ij}^{D}}{\sqrt{%
2m_{d_{i}}m_{d_{j}}}}v\gamma_{5}$ \\ \hline
\end{tabular}%
\end{center}
\par
\vspace{0.25cm}
\caption{\textit{Yukawa couplings structure for 2HDM III (neutral Higgs with
quarks and charged leptons) in the fundamental parametrization. $\protect%
\alpha$ is the mixing angle between neutral gauge eigenstates and mass
eigenstates (CP-even) \protect\cite{Diaz}.}}
\label{tabla de acoples down I}
\end{table}

By using the Cheng-Sher parametrization Eq. (\ref{cheng-sher}) for diagonal
couplings, we obtain the unitarity constraints by combining Eqs. (\ref{uni}%
), (\ref{coefi}) and (\ref{a01'}). They are given by

\begin{equation}
|\lambda _{ii}^{U,D}|\leq \left( \frac{2\sqrt{{2}}{\pi }}{G_{f}m_{f}^{2}}-%
\frac{{1}}{2}\right) ^{1/2}.  \label{acoFCNCuni}
\end{equation}

These relations lead to upper bounds for the fermion generations. We obtain
them by taking the input parameters in \cite{PDG}, and they are specified in
the caption of table \ref{tab:acoplesyukawa1}.

\begin{table}[tph]
\centering
\par
\begin{center}
\begin{tabular}{|c|c|c|}
\hline
$\lambda_{ii}^{U,D}$ & $\vert \lambda_{ii}\vert_{unit}$ & $\sim\mathcal{O}$
\\ \hline\hline
$\lambda_{tt}$ & 5 & $\mathcal{O}(1)$ \\ \hline
$\lambda_{bb}$ & 208 & $\mathcal{O}(10^{2})$ \\ \hline
$\lambda_{cc}$ & 687 & $\mathcal{O}(10^{2})$ \\ \hline
$\lambda_{ss}$ & 8.6 $\times10^{3}$ & $\mathcal{O}(10^{3})$ \\ \hline
$\lambda_{uu}$ & (2.6-5.1) $\times10^{5}$ & $\mathcal{O}(10^{5})$ \\ \hline
$\lambda_{dd}$ & (1.5-2.1) $\times10^{5}$ & $\mathcal{O}(10^{5})$ \\ \hline
$\lambda_{\tau\tau}$ & 491 & $\mathcal{O}(10^{2})$ \\ \hline
$\lambda_{\mu\mu}$ & 8.3 $\times10^{3}$ & $\mathcal{O}(10^{3})$ \\ \hline
$\lambda_{ee}$ & 1.7 $\times10^{6}$ & $\mathcal{O}(10^{6})$ \\ \hline
\end{tabular}%
\end{center}
\par
\label{tab:acoplesdeyukawanomezcla} \vspace{0.25cm}
\caption{\textit{Bounds on Yukawa couplings from relation (\protect\ref%
{acoFCNCuni}) for elastic processes ${f}_{i}\bar{f}_{i}\rightarrow f_{i}\bar{%
f}_{i}$ in the 2HDM type III. The parameters were taken from \protect\cite%
{PDG} (central values): $m_{t}=172$ GeV, $m_{b}=4.19$ GeV, $m_{c}=1.27$ GeV, 
$m_{s}=0.101$ GeV, $m_{d}=(0.0041-0.0058)$ GeV, $m_{u}=(0.0017-0.0033)$ GeV, 
$m_{\protect\tau }=1.776$ GeV, $m_{\protect\mu }=0.106$ GeV and $%
m_{e}=0.00051$ GeV.}}
\label{tab:acoplesyukawa1}
\end{table}

In the same way, for non-diagonal couplings, the upper bounds become

\begin{equation}
|\lambda _{ij}^{U,D}|\leq \left( \frac{2\sqrt{{2}}{\pi }}{%
G_{f}m_{_{i}}m_{_{j}}}\right) ^{1/2},  \label{acoFCNCunimix}
\end{equation}

and they are evaluated numerically in table \ref{tab:acoplesyukawa}.

\begin{table}[tph]
\centering
\par
\begin{center}
\begin{tabular}{|c|c|c|}
\hline
$\lambda_{ij}^{U,D}$ & $\vert \lambda_{ij}\vert_{unit}$ & $\sim\mathcal{O}$
\\ \hline\hline
$\lambda_{tc}$ & $59$ & $\mathcal{O}(10)$ \\ \hline
$\lambda_{tu}$ & $1.2-1.6\times10^{3}$ & $\mathcal{O}(10^{3})$ \\ \hline
$\lambda_{bs}$ & $1.3 \times 10^{3}$ & $\mathcal{O}(10^{3})$ \\ \hline
$\lambda_{bd}$ & $5.6-6.6\times10^{3}$ & $\mathcal{O}(10^{3})$ \\ \hline
$\lambda_{\tau\mu}$ & $2.0\times10^{3}$ & $\mathcal{O}(10^{3})$ \\ \hline
$\lambda_{\tau e}$ & $2.9\times10^{4}$ & $\mathcal{O}(10^{4})$ \\ \hline
$\lambda_{\mu e}$ & $1.2\times10^{5}$ & $\mathcal{O}(10^{5})$ \\ \hline
\end{tabular}%
\end{center}
\par
\vspace{0.25cm}
\caption{\textit{Bounds on Yukawa couplings from (\protect\ref{acoFCNCunimix}%
) for mixed channels ${f}_{i}\bar{f}_{j}\rightarrow f_{i}\bar{f}_{j}$ in the
2HDM type III. The parameters were taken from \protect\cite{PDG}, and they
are specified in the caption of table \protect\ref%
{tab:acoplesdeyukawanomezcla}.}}
\label{tab:acoplesyukawa}
\end{table}

All unitarity constraints compete with those coming from perturbativity, in
which we require that the running coupling constants of the Higgs
self-couplings and the Yukawa couplings do not blow up below a certain
energy scale $\Lambda $ : $\lambda _{i}(\mu )<8\pi $ and $(g_{f}^{\eta }(\mu
))^{2}<4\pi $, for a renormalization scale $\mu $ less than $\Lambda $ \cite%
{Kanemurapert}\footnote{%
In particular, from pseudoscalar-fermion couplings perturbativity requires
that: $\xi _{ij}^{2}<4\pi $. It is is translated into $\lambda _{ij}<\left( 
\frac{\sqrt{2}\pi }{G_{f}m_{i}m_{j}}\right) ^{1/2}$. Since the scalar
couplings depend on the mixing angle $\alpha $ as well as on the elements $%
\lambda _{ij}$, the perturbative constraints depend on more degrees of
freedom.}. If these couplings were higher, their $\beta $-functions will be
positive and their renormalization scale evolution will drive them to even
higher values \cite{Perturbativity}.

With mixed channels (e.g. $f_{i}\bar{f_{i}}\rightarrow f_{j}\bar{{f_{j}}}$
), the partial wave coefficients (\ref{coefi}) are transformed into

\begin{subequations}
\begin{align}
a^{0}(f_{\uparrow}\bar{f}_{\uparrow}\to f_{\uparrow}\bar{f}_{\uparrow}) & =-%
\frac{{\sqrt{2}G_{f}m_{i}m_{j}}}{16\pi}\sum_{\eta_{all}^{0}}\left(%
\Xi_{f_{i}f_{i}}^{\eta^{0}}\Xi_{f_{j}f_{j}}^{\eta^{0}}\right), \\
a^{0}(f_{\downarrow}\bar{f}_{\downarrow}\to f_{\uparrow}\bar{f}_{\uparrow})
& =-\frac{{\sqrt{2}G_{f}m_{i}m_{j}}}{16\pi}\sum_{\eta_{all}^{0}}\left[%
\Xi_{f_{i}f_{i}}^{\eta^{0}}\Xi_{f_{j}f_{j}}^{\eta^{0}}-\left(%
\Xi_{f_{i}f_{j}}^{\eta^{0}}\right)^{2}\right].  \label{a0mixed1}
\end{align}

From which the unitarity limits become

\end{subequations}
\begin{subequations}
\begin{align}
\sum_{\eta^{0}} \Xi^{\eta^{0}}_{f_{i}f_{i}}\Xi^{\eta^0}_{f_{j}f_{j}}\leq 
\frac{8\sqrt{2}\pi}{G_{f}m_{i}m_{j}}, \\
\sum_{\eta^{0}}\left[\Xi^{\eta^0}_{f_{i}f_{i}}\Xi^{\eta^0}_{f_{j}f_{j}}-%
\left(\Xi^{\eta^{0}}_{f_{i}f_{j}}\right)^{2}\right]\leq \frac{8\sqrt{2}\pi}{%
G_{f}m_{i}m_{j}}.
\end{align}

The sum runs over all neutral Higgs states (CP-even and CP-odd). In the
particular case of the 2HDM type III (from couplings in table \ref{tabla de
acoples down I}), these relations in terms of Sher-Cheng couplings satisfy

\end{subequations}
\begin{subequations}
\begin{align}
\lambda_{ii}^{U,D}\lambda_{jj}^{U,D}&\leq \frac{4\sqrt{2}\pi}{G_{f}m_{i}m_{j}%
}-\frac{1}{2},  \label{mixed1} \\
\lambda_{ii}^{U,D}\lambda_{jj}^{U,D}-\left(\lambda_{ij}^{U,D}\right)^{2}&%
\leq \frac{4\sqrt{2}\pi}{G_{f}m_{i}m_{j}}-\frac{1}{2} .  \label{mixed2}
\end{align}


Where the products are only by pairs either $\lambda _{ii}^{U}\lambda
_{jj}^{U}$ or $\lambda _{ii}^{D}\lambda _{jj}^{D}$. The numerical
evaluations of these crossed products for FCNC couplings are displayed in
table \ref{tab:mixedall}. It worths saying that if some of these couplings were
determined, these bounds could help in restricting the remaining ones.

\begin{table}[tph]
\centering
\par
\begin{center}
\begin{tabular}{|c|c|c|c|}
\hline
$\lambda_{ii}^{U,D}\lambda_{jj}^{U,D}$ & $\lambda_{ii}^{U,D}%
\lambda_{jj}^{U,D}-\left(\lambda_{ij}^{U,D}\right)^{2}$ & Bound & $\sim%
\mathcal{O}$ \\ \hline\hline
$\lambda_{tt}\lambda_{cc}$ & $\lambda_{tt}\lambda_{cc}-\lambda_{tc}^{2}$ & 
7.0$\times10^{3}$ & $\mathcal{O}(10^{3})$ \\ \hline
$\lambda_{tt}\lambda_{uu}$ & $\lambda_{tt}\lambda_{uu}-\lambda_{tu}^{2}$ & 
(2.7-5.2)$\times 10^{6}$ & $\mathcal{O}(10^{6})$ \\ \hline
$\lambda_{bb}\lambda_{ss}$ & $\lambda_{bb}\lambda_{ss}-\lambda_{bs}^{2}$ & 
3.6$\times 10^{6}$ & $\mathcal{O}(10^{6})$ \\ \hline
$\lambda_{bb}\lambda_{dd}$ & $\lambda_{bb}\lambda_{dd}-\lambda_{bd}^{2}$ & 
(6.3-8.8)$\times 10^{7}$ & $\mathcal{O}(10^{7})$ \\ \hline
$\lambda_{\tau\tau}\lambda_{\mu\mu}$ & $\lambda_{\tau\tau}\lambda_{\mu\mu}-%
\lambda_{\tau\mu}^{2}$ & 8.2$\times10^{6}$ & $\mathcal{O}(10^{6})$ \\ \hline
$\lambda_{\tau\tau}\lambda_{ee}$ & $\lambda_{\tau\tau}\lambda_{ee}-\lambda_{%
\tau e}^{2}$ & 1.7$\times10^{9}$ & $\mathcal{O}(10^{9})$ \\ \hline
$\lambda_{\tau\tau}\lambda_{ee}$ & $\lambda_{\tau\tau}\lambda_{ee}-\lambda_{%
\tau e}^{2}$ & 2.8$\times10^{10}$ & $\mathcal{O}(10^{10})$ \\ \hline
\end{tabular}%
\end{center}
\par
\vspace{0.25cm}
\label{tab:mixedall}
\caption{\textit{Bounds on Yukawa couplings from (\protect\ref{mixed1}) and (%
\protect\ref{mixed2}) for mixed channels ${f}_{i}\bar{f}_{i}\rightarrow f_{j}%
\bar{f}_{j}$ in the 2HDM type III. The parameters was taken from 
\protect\cite{PDG}, and they are specified in table \protect\ref%
{tab:acoplesdeyukawanomezcla}.}}
\end{table}

In the same way in which the unitarity constraints are interpreted for
self-couplings of the Higgs potential, these Yukawa couplings constraints
can be treated (without inclusion of new physics) as the upper values for
which the perturbation theory will become reliable at all energy scales.

\section{Charged Channels. \label{charged}}

It is also possible to evaluate the contribution from charged channels
(final and initial charged states) to the unitary amplitude. It is
worthwhile to observe that the matrix elements $\xi _{ij}$ modify the
charged Higgs couplings: 
\end{subequations}
\begin{equation*}
\chi _{f_{i}f_{j}}^{H^{\pm }}=(K_{ik}\xi _{kj}^{D}P_{R}-\xi
_{ik}^{U}K_{kj}P_{L}),
\end{equation*}%
for the fundamental parametrization \cite{Diaz,Sher}, where $K$ is the
Kobayashi Maskawa matrix and $P_{L(R)}$ are the Left(Right) projection
operators. Hence, there are two facts to point out i) the flavor changing
charged currents (FCCC) in the quark sector are modified by the same matrix
that produces FCNC, ii) in the lepton sector FCCC are generated by the same
matrix that generates FCNC.

A typical charged scattering process at the tree level for the scalar sector
has two contributions: the first one associated with $H^{\pm }$ states in
the propagator for the $s$-channel and the second one with neutral scalar
states in the propagator for the $t$-channel:

\begin{equation*}
\mathcal{M}(f_{i}\bar{f}_{j}\rightarrow f_{i}\bar{f}_{j})=\sum_{\eta
_{H^{\pm }}}\left( \bar{v}_{2}\chi _{f_{i}f_{j}}^{H^{\pm }}u_{1}\frac{1}{%
s-m_{\eta _{H^{\pm }}}^{2}}\bar{u}_{3}\chi _{f_{i}f_{j}}^{H^{\pm
}}v_{4}\right) +\sum_{\eta ^{0}}\left( \bar{v}_{2}i\chi _{f_{j}f_{j}}^{\eta
^{0}}v_{4}\frac{1}{t-m_{\eta ^{0}}^{2}}\bar{u}_{3}i\chi _{f_{i}f_{i}}^{\eta
^{0}}u_{1}\right) .
\end{equation*}

Assuming diagonal textures for both flavor matrices in 2HDM type III (quark
sector) and using the systematic got in the last section, the amplitudes for
the polarized process $f_{i\uparrow }\bar{f}_{j\uparrow }\rightarrow
f_{i\uparrow }\bar{f}_{j\uparrow }$ becomes

\begin{align}
\mathcal{M}(f_{i\uparrow}\bar{f}_{j\uparrow}\to f_{i\uparrow}\bar{f}%
_{j\uparrow})=\bar{v}_{2_{\uparrow}}\chi^{H^{\pm}}_{f_{i}f_{j}}u_{1_{%
\uparrow}}\frac{1}{s-m_{H^{\pm}}^{2}}\bar{u}_{3_{\uparrow}}\chi^{H^{\pm
}}_{f_{i}f_{j}}v_{4_{\uparrow}}=\xi_{ii}^{U}\xi_{jj}^{D}K_{ij}^{2}\frac{s}{%
s-m_{H^{\pm}}^{2}}.
\end{align}

where $i=u,c,t$ and $j=d,s,b$. We have used the CKM hierarchy and the
assumption of universality deviation in the same generation. Here $%
u_{1\uparrow },\bar{v}_{2\uparrow },\bar{u}_{3\uparrow }$ and $v_{4\uparrow
} $ are the eigenspinors as (right-handed) helicity states. Since $\bar{u}%
_{3_{\uparrow }}u_{1_{\uparrow }}=0$ at the high energy limit (appendix \ref%
{appendix}), the neutral channel does not have a contribution for this
helicity choice. From Cheng-Sher anzats for diagonal couplings, the partial
wave coefficient has the unitarity bound

\begin{align}  \label{charged1}
\lambda_{ii}^{U}\lambda_{jj}^{D}\leq \frac{2\sqrt{2}\pi}{%
G_{f}m_{i}m_{j}K_{ij}^{2}}.
\end{align}

We have summarized these bounds in the table \ref{tab:charged}.

If some of these couplings were determined (say couplings from the up
sector), these bounds could help in restricting the remaining ones (say
couplings from the down sector)\footnote{%
For instance, the process $b\rightarrow s\gamma $ could determine the value
of $\lambda _{tt}$, from which our present bounds would help in obtaining
the associated $\lambda _{dd}$.}.

\begin{table}[tph]
\centering
\par
\begin{center}
\begin{tabular}{|c|c|c|}
\hline
$\lambda_{ij}^{U}\lambda_{jj}^{D}$ & Bound & $\sim\mathcal{O}$ \\ 
\hline\hline
$\lambda_{tt}\lambda_{bb}$ & $1055$ & $\mathcal{O}(10^{3})$ \\ \hline
$\lambda_{cc}\lambda_{ss}$ & $5.9\times10^{6}$ & $\mathcal{O}(10^{6})$ \\ 
\hline
$\lambda_{uu}\lambda_{dd}$ & (0.4-1.1) $\times 10^{11}$ & $\mathcal{O}%
(10^{3})$ \\ \hline
\end{tabular}%
\end{center}
\par
\vspace{0.25cm}
\caption{\textit{Bounds on Yukawa couplings from (\protect\ref{charged1})
for mixed channels ${f}_{i}\bar{f}_{j}\rightarrow f_{i}\bar{f}_{j}$ in the
2HDM type III. The parameters was taken from \protect\cite{PDG}. and they
are specified in table \protect\ref{tab:acoplesdeyukawanomezcla}.}}
\label{tab:charged}
\end{table}

%
%
%
%
%
%

\section{Conclusions and Remarks}

Under the use of a general expansion of partial waves, unitarity constraints
over fermionic scattering processes were obtained. The method relies on a
diagonalization (at least partial) in the angular momentum basis of the $%
\widehat{S}$ matrix for the scattering of two particles in the center of
mass frame and the appropriate choice of helicity states. In the helicity
formalism, the spin degrees of freedom of the particle involved do not
introduce any significant complication with respect to spinless particles,
at least when $\lambda _{a}=\lambda _{b}$ (initial helicities) and $\lambda
_{c}=\lambda _{d}$ (final helicities). In fact, this particular case
recovers the traditional partial wave expansion as well as its unitary
conditions over the coefficients expansion. This leads to build up a well
grounded formalism in order to impose the unitary constraints over spin $1/2$
states or in general states of any spin.

Due to the universality deviation by the presence of FCNC vertices for
fermionic interactions with scalars in the type III 2HDM, this formalism is
applied in all its generality to get unitary constraints over Yukawa
couplings values under Cheng and Sher parametrization. The constraints
obtained are indeed independent of other parameters of the Higgs sector,
i.e. the scalar masses and mixing angles. Such constraints only depend on
the fermionic masses which are input parameters.

In the case of elastic scattering processes, diagonal Yukawa couplings
constraints (coming from $f_{i}\bar{f}_{i}\rightarrow f_{i}\bar{f}_{i}$) are
more stringent that non-diagonal couplings (coming from $f_{i}\bar{f}%
_{k}\rightarrow f_{i}\bar{f}_{k}$). In addition, these unitarity limits
compete with those imposed from perturbative interactions, which could
introduce more parameters. Moreover, our limitations compete with current
phenomenological constraints ($\bar{B}^{0}-B^{0}$ mixing, $(g-2)_{\mu }$
factor) for heavy fermionic masses, e.g. the top mass. It is worthwhile
emphasizing that the phenomenological constraints demand the use of several
parameters like scalar masses or mixing angles. Further, those computations
lie on two-loop radiative corrections to the physical processes unlike the
unitary constraints, which unfold naturally at the tree level.

Finally, this systematic might be extrapolated to other fermionic sectors
such as the minimal supersymmetric standard model, minimal $B-L$ extension
of the SM, the SM with fourth generations, etc.

\appendix

\section{Computations of the polarized amplitudes}

\label{appendix}

We consider scattering amplitudes of the type $f\bar{f}\rightarrow f\bar{f}$
by means of the helicity formalism when $s\gg m_{\eta ^{0}}^{2}$ and $%
m_{\eta ^{0}}^{2}\gg m_{z}^{2},m_{w}^{2}$. These developments are useful in
the fermionic unitarity constraints. In that regime, Feynman diagrams for
these processes are displayed in Fig. \ref{fig:ff}. The matrix element for
CP-even states is

%
%
%
%
%
%
%

\begin{align}
\mathcal{M}(f\bar{f}\to f \bar{f})=-\sqrt{2}G_{f}m_{f}^{2}\sum_{\eta^{0}}%
\left(\Xi_{f\bar{f}}^{\eta^{0}}\right)^{2}\left(\bar{v}_{2}u_{1}\frac{1}{%
s-m_{\eta^{0}}^{2}}\bar{u}_{3}v_{4}+\bar{v}_{2}v_{4}\frac{1}{%
t-m_{\eta^{0}}^{2}}\bar{u}_{3}u_{1}\right).
\end{align}

Here the relative couplings with respect to the SM ones ($\Xi _{f\bar{f}%
}^{\eta ^{0}}$) have been factorized. The eigenspinors as helicity states in
spherical polar coordinates have the form \cite{Thomson}:

\begin{subequations}
\begin{align}
u_{\uparrow}(p)=\sqrt{E+m}%
\begin{pmatrix}
\cos\frac{\theta}{2} \\ 
e^{i\phi}\sin\frac{\theta}{2} \\ 
\frac{\vert \mathbf{p}\vert}{E+m}\cos\frac{\theta}{2} \\ 
\frac{\vert \mathbf{p}\vert}{E+m}e^{i\phi}\sin\frac{\theta}{2}%
\end{pmatrix}%
, \hspace{0.3cm} u_{\downarrow}(p)=\sqrt{E+m}%
\begin{pmatrix}
-\sin\frac{\theta}{2} \\ 
e^{i\phi}\cos\frac{\theta}{2} \\ 
\frac{\vert \mathbf{p}\vert}{E+m}\sin\frac{\theta}{2} \\ 
-\frac{\vert \mathbf{p}\vert}{E+m}e^{i\phi}\cos\frac{\theta}{2}%
\end{pmatrix}%
, \\
v_{\uparrow}(p)=\sqrt{E+m}%
\begin{pmatrix}
\frac{\vert \mathbf{p}\vert}{E+m}\sin\frac{\theta}{2} \\ 
-\frac{\vert \mathbf{p}\vert}{E+m}e^{i\phi}\cos\frac{\theta}{2} \\ 
-\sin\frac{\theta}{2} \\ 
e^{i\phi}\cos\frac{\theta}{2}%
\end{pmatrix}%
, \hspace{0.3cm} v_{\downarrow}(p)=\sqrt{E+m}%
\begin{pmatrix}
\frac{\vert \mathbf{p}\vert}{E+m}\cos\frac{\theta}{2} \\ 
\frac{\vert \mathbf{p}\vert}{E+m}e^{i\phi}\sin\frac{\theta}{2} \\ 
\cos\frac{\theta}{2} \\ 
e^{i\phi}\sin\frac{\theta}{2}%
\end{pmatrix}%
.
\end{align}
At the high energy limit $E\gg m$, we obtain

\end{subequations}
\begin{subequations}
\begin{align}
u_{\uparrow}(p)=\sqrt{E}%
\begin{pmatrix}
\cos\frac{\theta}{2} \\ 
e^{i\phi}\sin\frac{\theta}{2} \\ 
\cos\frac{\theta}{2} \\ 
e^{i\phi}\sin\frac{\theta}{2}%
\end{pmatrix}%
, \hspace{0.3cm} u_{\downarrow}(p)=\sqrt{E}%
\begin{pmatrix}
-\sin\frac{\theta}{2} \\ 
e^{i\phi}\cos\frac{\theta}{2} \\ 
\sin\frac{\theta}{2} \\ 
-e^{i\phi}\cos\frac{\theta}{2}%
\end{pmatrix}%
, \\
v_{\uparrow}(p)=\sqrt{E}%
\begin{pmatrix}
\sin\frac{\theta}{2} \\ 
-e^{i\phi}\cos\frac{\theta}{2} \\ 
-\sin\frac{\theta}{2} \\ 
e^{i\phi}\cos\frac{\theta}{2}%
\end{pmatrix}%
, \hspace{0.3cm} v_{\downarrow}(p)=\sqrt{E}%
\begin{pmatrix}
\cos\frac{\theta}{2} \\ 
e^{i\phi}\sin\frac{\theta}{2} \\ 
\cos\frac{\theta}{2} \\ 
e^{i\phi}\sin\frac{\theta}{2}%
\end{pmatrix}%
.
\end{align}

Moreover in the center of mass frame, the 4-momenta can be written as

\end{subequations}
\begin{subequations}
\begin{align}
p_{1}&=\left(E,0,0,p\right); \hspace{0.3cm} p_{2}=\left(E,0,0,-p\right). \\
p_{3}&=\left(E,\overrightarrow{p}_{f}\right); \hspace{0.3cm} p_{4}=\left(E,-%
\overrightarrow{p}_{f}\right).
\end{align}

Where $E\equiv E_{cm}/2$. Figure \ref{fig:kinematic} displays the kinematic
parametrization used for the eigenspinors evaluation.

\begin{figure}[tph]
\centering\includegraphics[scale=0.32]{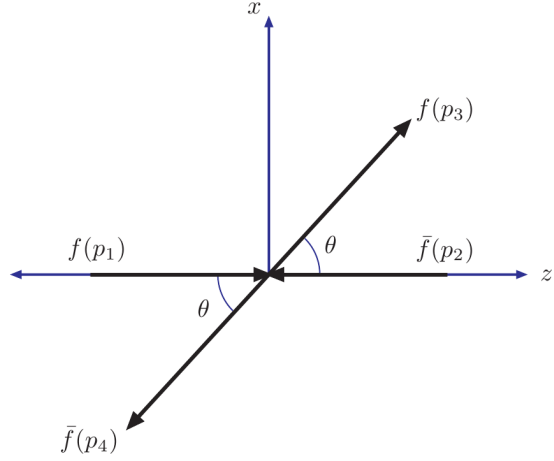} \vspace{0.4cm}
\caption{\textit{Kinematic parametrization used for the scattering process $f%
\bar{f}\rightarrow f\bar{f}$ in the center of mass system.}}
\label{fig:kinematic}
\end{figure}

The initial state of the fermion can be of up or down helicity. By taking $%
\phi =0,\ $it reads

\end{subequations}
\begin{align}
u_{\uparrow}(p_{1})=\sqrt{E}%
\begin{pmatrix}
1 \\ 
0 \\ 
1 \\ 
0%
\end{pmatrix}%
; \hspace{0.3cm} u_{\downarrow}(p_{1})=\sqrt{E}%
\begin{pmatrix}
0 \\ 
1 \\ 
0 \\ 
-1%
\end{pmatrix}%
.
\end{align}

For antifermion initial states ($\theta =\pi $) we have

\begin{align}
v_{\uparrow}(p_{2})=\sqrt{E}%
\begin{pmatrix}
1 \\ 
0 \\ 
-1 \\ 
0%
\end{pmatrix}%
; \hspace{0.3cm} v_{\downarrow}(p_{2})=\sqrt{E}%
\begin{pmatrix}
0 \\ 
1 \\ 
0 \\ 
1%
\end{pmatrix}%
.
\end{align}

Hence, the final fermion states, (choosing $\phi =0$) yield

\begin{align}
u_{\uparrow}(p)=\sqrt{E}%
\begin{pmatrix}
\cos\frac{\theta}{2} \\ 
\sin\frac{\theta}{2} \\ 
\cos\frac{\theta}{2} \\ 
\sin\frac{\theta}{2}%
\end{pmatrix}%
, \hspace{0.3cm} u_{\downarrow}(p)=\sqrt{E}%
\begin{pmatrix}
-\sin\frac{\theta}{2} \\ 
\cos\frac{\theta}{2} \\ 
\sin\frac{\theta}{2} \\ 
-\cos\frac{\theta}{2}%
\end{pmatrix}%
,
\end{align}

while for antifermions ($\theta \rightarrow \pi -\theta $ and $\phi =\pi $)
we obtain

\begin{align}
v_{\uparrow}(p)=\sqrt{E}%
\begin{pmatrix}
\cos\frac{\theta}{2} \\ 
\sin\frac{\theta}{2} \\ 
-\cos\frac{\theta}{2} \\ 
-\sin\frac{\theta}{2}%
\end{pmatrix}%
, \hspace{0.3cm} v_{\downarrow}(p)=\sqrt{E}%
\begin{pmatrix}
\sin\frac{\theta}{2} \\ 
-\cos\frac{\theta}{2} \\ 
\sin\frac{\theta}{2} \\ 
-\cos\frac{\theta}{2}%
\end{pmatrix}%
.
\end{align}

For instance for $f_{\uparrow}\bar{f}_{\uparrow}\to f_{\uparrow}\bar{f}%
_{\uparrow}$, the matrix element is

\begin{align}  \label{am1}
\mathcal{M}(f_{\uparrow}\bar{f}_{\uparrow}\to f_{\uparrow} \bar{f}%
_{\uparrow})=-\sqrt{2}G_{f}m_{f}^{2}\sum_{\eta^{0}}\left(\Xi_{f\bar{f}%
}^{\eta^{0}}\right)^{2}\left(\bar{v}_{2\uparrow}u_{1\uparrow}\frac{1}{%
s-m_{\eta^{0}}^{2}}\bar{u}_{3\uparrow}v_{4\uparrow}+\bar{v}%
_{2\uparrow}v_{4\uparrow}\frac{1}{t-m_{\eta^{0}}^{2}} \bar{u}%
_{3\uparrow}u_{1\uparrow}\right).
\end{align}

calculating all the products for initial states we have

\begin{align}
\bar{v}_{2\uparrow
}u_{1\uparrow}=v_{2\uparrow}^{\dagger}\gamma^{0}u_{1\uparrow}&=E%
\begin{pmatrix}
1,0,-1,0%
\end{pmatrix}
\begin{pmatrix}
1 & 0 & 0 & 0 \\ 
0 & 1 & 0 & 0 \\ 
0 & 0 & -1 & 0 \\ 
0 & 0 & 0 & -1%
\end{pmatrix}
\begin{pmatrix}
1 \\ 
0 \\ 
1 \\ 
0%
\end{pmatrix}%
,  \notag \\
&=E%
\begin{pmatrix}
1,0,1,0%
\end{pmatrix}
\begin{pmatrix}
1 \\ 
0 \\ 
1 \\ 
0%
\end{pmatrix}%
=2E.
\end{align}

while for final states they are given by

\begin{align}
\bar{u}_{3\uparrow}v_{4\uparrow}=u_{3\uparrow}^{\dagger}\gamma^{0}v_{4%
\uparrow}&=E%
\begin{pmatrix}
\cos\frac{\theta}{2}, \sin\frac{\theta}{2},\cos\frac{\theta}{2},\sin\frac{%
\theta}{2}%
\end{pmatrix}
\begin{pmatrix}
1 & 0 & 0 & 0 \\ 
0 & 1 & 0 & 0 \\ 
0 & 0 & -1 & 0 \\ 
0 & 0 & 0 & -1%
\end{pmatrix}
\begin{pmatrix}
\cos\frac{\theta}{2} \\ 
\sin\frac{\theta}{2} \\ 
-\cos\frac{\theta}{2} \\ 
-\sin\frac{\theta}{2}%
\end{pmatrix}%
=2E.
\end{align}

In the same way, for the mixing between antifermions in initial and final
states we have

\begin{align}
\bar{v}_{2\uparrow}v_{4\uparrow}=v_{2}^{\dagger}\gamma^{0}v_{4}&=E%
\begin{pmatrix}
1,0,-1,0%
\end{pmatrix}%
\begin{pmatrix}
1 & 0 & 0 & 0 \\ 
0 & 1 & 0 & 0 \\ 
0 & 0 & -1 & 0 \\ 
0 & 0 & 0 & -1%
\end{pmatrix}
\begin{pmatrix}
\cos\frac{\theta}{2} \\ 
\sin\frac{\theta}{2} \\ 
-\cos\frac{\theta}{2} \\ 
-\sin\frac{\theta}{2}%
\end{pmatrix}%
=0.
\end{align}

For the mixing between fermions in their initial and final states we obtain

\begin{align}
\bar{u}_{3\uparrow}u_{1\uparrow}=u_{3\uparrow}^{\dagger}\gamma^{0}u_{1%
\uparrow}&=E%
\begin{pmatrix}
\cos\frac{\theta}{2}, \sin\frac{\theta}{2},\cos\frac{\theta}{2},\sin\frac{%
\theta}{2}%
\end{pmatrix}%
\begin{pmatrix}
1 & 0 & 0 & 0 \\ 
0 & 1 & 0 & 0 \\ 
0 & 0 & -1 & 0 \\ 
0 & 0 & 0 & -1%
\end{pmatrix}
\begin{pmatrix}
1 \\ 
0 \\ 
1 \\ 
0%
\end{pmatrix}%
=0.
\end{align}

Therefore, the matrix element is written by ($s=4E^{2}$)

\begin{align}  \label{Miii}
\mathcal{M}(f_{\uparrow}\bar{f}_{\uparrow}\to f_{\uparrow} \bar{f}%
_{\uparrow})=-\sqrt{2}G_{f}m_{f}^{2}\sum_{\eta^{0}}\left(\Xi_{f\bar{f}%
}^{\eta^{0}}\right)^{2}\left(\frac{s}{s-m_{\eta^{0}}^{2}}\right).
\end{align}

With a similar procedure it is shown that

\begin{align}
\mathcal{M}(f_{\downarrow}\bar{f}_{\downarrow}\to f_{\downarrow} \bar{f}%
_{\downarrow})=\sqrt{2}G_{f}m_{f}^{2}\sum_{\eta^{0}}\left(\Xi_{f\bar{f}%
}^{\eta^{0}}\right)^{2}\left(\frac{s}{s-m_{\eta^{0}}^{2}}\right).
\end{align}

Now, we consider a mixed channel in helicity states for scattering amplitude
with the structure

\begin{align}
\mathcal{M}(f_{\downarrow}\bar{f}_{\downarrow}\to f_{\uparrow} \bar{f}%
_{\uparrow})=-\sqrt{2}G_{f}^{2}m_{f}^{2}\sum_{\eta^{0}}\left(\Xi_{f\bar{f}%
}^{\eta^{0}}\right)^{2}\left(\bar{v}_{2\downarrow}u_{1\downarrow}\frac{1}{%
s-m_{\eta^{0}}^{2}}\bar{u}_{3\uparrow}v_{4\uparrow}+\bar{v}%
_{2\downarrow}v_{4\uparrow}\frac{1}{t-m_{\eta^{0}}^{2}}\bar{u}%
_{3\uparrow}u_{1\downarrow}\right).
\end{align}

The products are the following

\begin{align}
\bar{v}_{2\downarrow}u_{1\downarrow}=v_{2\downarrow}^{\dagger}\gamma^{0}u_{1%
\downarrow}&=E%
\begin{pmatrix}
0,1,0,1%
\end{pmatrix}%
\begin{pmatrix}
1 & 0 & 0 & 0 \\ 
0 & 1 & 0 & 0 \\ 
0 & 0 & -1 & 0 \\ 
0 & 0 & 0 & -1%
\end{pmatrix}
\begin{pmatrix}
0 \\ 
1 \\ 
0 \\ 
-1%
\end{pmatrix}%
=2E.
\end{align}

\begin{align}
\bar{v}_{2\downarrow}v_{4\uparrow}={v}_{2\downarrow}^{\dagger}\gamma^{0}v_{4%
\uparrow}&=E%
\begin{pmatrix}
0,1,0,1%
\end{pmatrix}%
\begin{pmatrix}
1 & 0 & 0 & 0 \\ 
0 & 1 & 0 & 0 \\ 
0 & 0 & -1 & 0 \\ 
0 & 0 & 0 & -1%
\end{pmatrix}
\begin{pmatrix}
\cos\frac{\theta}{2} \\ 
\sin\frac{\theta}{2} \\ 
-\cos\frac{\theta}{2} \\ 
-\sin\frac{\theta}{2}%
\end{pmatrix}%
=2E\sin\frac{\theta}{2}.
\end{align}

Since $\bar{u}_{3\uparrow }v_{4\uparrow }=2E$ and using

\begin{align}
\bar{u}_{3\uparrow}u_{1\downarrow}=u_{3\uparrow}^{\dagger}\gamma^{0}u_{1%
\downarrow}&=E%
\begin{pmatrix}
\cos\frac{\theta}{2}, \sin\frac{\theta}{2},\cos\frac{\theta}{2},\sin\frac{%
\theta}{2}%
\end{pmatrix}%
\begin{pmatrix}
1 & 0 & 0 & 0 \\ 
0 & 1 & 0 & 0 \\ 
0 & 0 & -1 & 0 \\ 
0 & 0 & 0 & -1%
\end{pmatrix}
\begin{pmatrix}
1 \\ 
0 \\ 
-1 \\ 
0%
\end{pmatrix}
=2E\sin\frac{\theta}{2},
\end{align}

the matrix element becomes

\begin{align}  \label{ii||}
\mathcal{M}(f_{\downarrow}\bar{f}_{\downarrow}\to f_{\uparrow} \bar{f}%
_{\uparrow})=-\sqrt{2}G_{f}m_{f}^{2}\sum_{\eta^{0}}\left(\Xi_{f\bar{f}%
}^{\eta^{0}}\right)^{2}\left(\frac{s}{s-m_{\eta^{0}}^{2}}+\frac{s}{2}\frac{%
(1-\cos\theta)}{t-m_{\eta^{0}}^{2}}\right).
\end{align}

In the frame displayed in Fig. \ref{fig:kinematic}, Mandelstam variables
(with $m_{f1}=m_{f2}$ y $m_{f3}=m_{f4}$\footnote{%
For any combination of initial states, the Mandelstam variable $s$ reads
\par
\begin{align*}
s& =(p_{1}+p_{2})^{2}=m_{f_{1}}^{2}+m_{f_{2}}^{2}+2E_{{1}{cm}}E_{2{cm}%
}+2E_{1cm}E_{2cm}\sqrt{1-\frac{m_{f_{2}}^{2}}{E_{2cm}^{2}}-\frac{%
m_{f_{1}}^{2}}{E_{1cm}^{2}}+\frac{m_{f_{1}}^{2}m_{f_{2}}^{2}}{E_{{1}{cm}}E_{2%
{cm}}}}.
\end{align*}%
\par
{}and at the high energy limit $s\rightarrow E_{cm}^{2}$.}) are described by

\begin{align}
s&=(p_{1}+p_{2})^{2}=2m_{f}^{2}+2p_{1}\cdot p_{2}=2m_{f}^{2}+2\frac{%
E_{cm}^{2}}{4}-2(\mathbf{p}_{1cm}\cdot\mathbf{p}_{2cm})=2m_{f}^{2}+\frac{%
E_{cm}^{2}}{2}-2\vert\mathbf{p}_{1cm}\vert^{2}\cos\theta_{12},  \notag \\
&=2m_{f}^{2}+\frac{E_{cm}^{2}}{2}+2\left(\frac{E_{cm}^{2}}{4}%
-m_{f}^{2}\right)=E_{cm}^{2}.  \label{s} \\
t&=(p_{1}-p_{3})=m_{f_{1}}^{2}+m_{f_{3}}^{2}-2p_{1}\cdot
p_{3}=m_{f_{1}}^{2}+m_{f_{3}}^{2}-2\left(\frac{E_{cm}^{2}}{4}-(\mathbf{p}%
_{1cm}\cdot\mathbf{p}_{3cm})\right),  \notag \\
&=m_{f_{1}}^{2}+m_{f_{3}}^{2}-2\left(\frac{E_{cm}^{2}}{4}-\sqrt{\left(\frac{%
E_{cm}^{2}}{4}-m_{f1}^{2}\right)\left(\frac{E_{cm}^{2}}{4}-m_{f3}^{2}\right)}%
\cos\theta\right) .  \label{t}
\end{align}

Replacing (\ref{s}) in (\ref{t}), we get

\begin{align}
t&=m_{f_{1}}^{2}+m_{f_{3}}^{2}-2\left(\frac{s}{4}-\sqrt{\left(\frac{s}{4}%
-m_{f1}^{2}\right)\left(\frac{s}{4}-m_{f_{3}}^{2}\right)}\cos\theta\right), 
\notag \\
&=m_{f_{1}}^{2}+m_{f_{3}}^{2}-\frac{s}{2}\left(1-\sqrt{1-\frac{4}{s}%
(m_{f1}^{2}+m_{f_{3}}^{2})+\frac{16}{s^{2}}m_{f_{1}}^{2}m_{f_{3}}^{2}}%
\cos\theta\right).
\end{align}

which for the high energy limit yields

\begin{align}  \label{t1}
t=-\frac{s}{2}\left(1-\cos\theta\right).
\end{align}

Taking into account (\ref{t1}) in the amplitude (\ref{ii||}), it is obtained

\begin{align}  \label{Miiii}
\mathcal{M}(f_{\downarrow}\bar{f}_{\downarrow}\to f_{\uparrow} \bar{f}%
_{\uparrow}) &=-\sqrt{2}G_{f}m_{f}^{2}\sum_{\eta^{0}}\left(\Xi_{f\bar{f}%
}^{\eta^{0}}\right)^{2}\left(\frac{s}{s-m_{\eta^{0}}^{2}}-\frac{t}{%
t-m_{\eta^{0}}^{2}}\right).
\end{align}

Under the same assumptions, it can be shown that the mixed channel becomes

\begin{align}
\mathcal{M}(f_{\uparrow}\bar{f}_{\uparrow}\to f_{\downarrow}\bar{f}%
_{\downarrow})=\sqrt{2}G_{f}m_{f}^{2}\sum_{\eta^{0}}\left(\Xi_{f\bar{f}%
}^{\eta^{0}}\right)^{2}\left(\frac{s}{s-m_{\eta^{0}}^{2}}-\frac{t}{%
t-m_{\eta^{0}}^{2}}\right).
\end{align}

Therefore we should consider only Eqs. (\ref{Miii}) and (\ref{Miiii}) as the
scattering amplitudes that emerge from a general treatment for the process $f%
\bar{f}\rightarrow f\bar{f}$ at the high energy limit and $m_{\eta
^{0}}>>m_{W},m_{z}$, with respect to the computed unitarity constraints on
free parameters in the scalar-fermionic sector of the 2HDM. Other channels
just differ by a minus sign. This fact is not relevant for the computation
of unitarity bounds since they depend on the $J=0$ partial wave norm.

\subsection*{Couplings with pseudoscalars $A^{0}$}

The amplitude with pseudoscalars couplings (CP-odd states) gives

\begin{align}
\mathcal{M}(f_{\uparrow}\bar{f}_{\uparrow}\to f_{\uparrow} \bar{f}%
_{\uparrow})=-\sqrt{2}G_{f}m_{f}^{2}\sum_{\eta^{0}_{CP-imp}}\left(\Xi_{f\bar{%
f}}^{\eta^{0}}\right)^{2}\left(i\bar{v}_{2\uparrow}\gamma_{5}u_{1\uparrow}%
\frac{1}{s-m_{\eta^{0}}^{2}}i\bar{u}_{3\uparrow}\gamma_{5}v_{4\uparrow}+i%
\bar{v}_{2\uparrow}\gamma_{5}v_{4\uparrow}\frac{1}{t-m_{\eta^{0}}^{2}} i\bar{%
u}_{3\uparrow}\gamma_{5}u_{1\uparrow}\right).
\end{align}

At the high energy limit $s\gg m^{2}$, the helicity states are also chiral
states. Then we have

\begin{align}
\gamma_{5}u_{\uparrow}=+u_{\uparrow}; \hspace{0.25cm} \gamma_{5}u_{%
\downarrow}=-u_{\downarrow}; \hspace{0.25cm} \gamma_{5}v_{\uparrow}=-v_{%
\uparrow}; \hspace{0.25cm} \gamma_{5}v_{\downarrow}=+v_{\downarrow}.
\end{align}

The products with $\gamma_{5}$ are the following

\begin{subequations}
\begin{align}
\bar{v}_{2\uparrow }\gamma _{5}u_{1\uparrow }& =\bar{v}_{2\uparrow
}u_{1\uparrow }\ \ \ ,\ \ \ \bar{u}_{3\uparrow }\gamma _{5}v_{4\uparrow }=-%
\bar{u}_{3\uparrow }v_{4\uparrow },  \label{amp1a} \\
\bar{v}_{2\uparrow }\gamma _{5}v_{4\uparrow }& =-\bar{v}_{2\uparrow
}v_{4\uparrow }\ \ \ ,\ \ \ \bar{u}_{3\uparrow }\gamma _{5}u_{1\uparrow }=%
\bar{u}_{3\uparrow }u_{1\uparrow }.  \label{amp1b}
\end{align}

Which reproduce the amplitude for scalars with CP-even states (\ref{am1}).
Another combinations for helicity states also satisfy these properties.
Hence, the scattering amplitude at the tree level with $s\gg m$ is the sum
among CP-even ones and CP-odd ones.

\end{subequations}

\end{document}